\documentclass[review, sort&compress]{elsarticle}

\usepackage{lineno,hyperref}
\usepackage{graphicx, epstopdf, amstext, amssymb,epsf,color,hyperref}

\modulolinenumbers[5]

\begin{document}

\begin{frontmatter}
\title{Evolution of popularity in given names}
\author[skku]{Mi Jin Lee}
\author[skku]{Woo Seong Jo}
\author[skku]{Il Gu Yi}
\author[pknu]{Seung Ki Baek\corref{cor1}}
\cortext[cor1]{seungki@pknu.ac.kr}
\author[skku]{Beom Jun Kim\corref{cor2}}
\cortext[cor2]{beomjun@skku.edu}
\address[skku]{Department of Physics, Sungkyunkwan University, Suwon 440-746, Korea}
\address[pknu]{Department of Physics, Pukyong National University, Busan 608-737, Korea}

\begin{abstract}
An individual's identity in a human society is specified by his or her
name. Differently from family names, usually inherited from fathers,
a given name for a child is often chosen at the parents' disposal.
However, their decision cannot be made in a vacuum but affected
by social conventions and trends. Furthermore, such social pressure
changes in time, as new names gain popularity while some other names
are gradually forgotten.
In this paper, we investigate how popularity of given names
has evolved over the last century by using datasets collected in Korea,
the province of Quebec in Canada, and the
United States. In each of these countries, the average popularity of given
names exhibits typical patterns of rise and fall with a time scale of
about one generation.
We also observe that notable changes of diversity in given names
signal major social changes.
\end{abstract}



\end{frontmatter}


\section{Introduction}
\label{sec:intro}

Since Galton statistically investigated extinction of families,
many researchers have studied dynamics of family names~\cite{usberlinfamily,jpfamily,master,smpark,kiet,*familymodel2,*familymodel3}.
The dynamics is well suited to mathematical analysis, because
family names are paternally inherited like the Y chromosome in most cases
[see, e.g., Ref.~\cite{rossi2013} for a review].
If we look at statistics of family names, the rank-size distribution
is broad in many countries~\cite{usberlinfamily,jpfamily},
whereas a clear exponential form is observed in Korea~\cite{master,smpark,kiet}.
These statistics can readily be explained by the branching process in
mathematics, and the essential ingredient to explain the Korean case
turns out to be a social taboo on changing family names~\cite{master}.
Although we have good mathematical understanding on its origin,
the exponential rank-size
distribution in Korea actually raises another question:
It has a characteristic rank scale beyond which minor family names are
found with very small frequencies.
Indeed, the top ten family names occupy roughly two
thirds of the total Korean population, which implies that it is virtually
impossible to identify individuals by using family names. Then, how do they
distinguish two different persons?
An obvious answer would be that the distinguishability is supplied by given
names~\footnote{
It does not mean that Koreans use a first-name basis in the daily life.
Rather, they have taboos on mentioning elders' given names.},
and one of our goals in this work is to examine whether this statement is
justified empirically.

Differently from family names, parents have a broad spectrum of
possible choices in picking up a given name, and the only criterion
is that it sounds good and proper. This is, however, rather subjective,
and what is worse is that the criterion itself changes
from generation to generation, and from one place to the other.
Nevertheless, researchers have tried to understand the given-name dynamics
by using empirical data~\cite{givenname1,*givenname2,*givenname3,*givenname4},
and a recent study suggests a typical temporal pattern of
rise and fall~\cite{fad}. However, the suggested pattern heavily relies on a
simplified model and no information is provided on its characteristic
time scale.
Therefore, we will verify the existence of such a pattern
and estimate its time scale on an empirical basis.

The rank-size distribution and temporal dynamics together determine
the diversity of names.
If this is directly related to distinguishability as argued above,
it will vary with the typical radius of social interactions: For
example, if one can live the whole life in a community consisting of
a small number of people, we would not need so many names.
In a modern society, however, the range of social interactions can be very
large, and it is no longer possible to define an individual in `relative'
coordinates like someone's son or someone's mother. Such a modern society is
sometimes called anonymous, but it is actually in this situation
that we expect the maximal diversity of names to distinguish every different
person. We note that
the social status of Korean women has drastically changed, with
increasing the radius of social interactions, over the last century.
Therefore, we hypothesize that Korean
female names have gradually become more and more diverse, compared to the
cases of the other countries, which will also be checked in this work.

The present paper is organized as follows: We introduce
the datasets used in this study in Sec.~\ref{sec:data}.
Two main results of the data analysis, i.e., the rank-size distribution and the
temporal pattern of popularity, are described in Sec.~\ref{subsec:ranksize}
and Sec.~\ref{subsec:riseandfall}, respectively. In Sec.~\ref{subsec:diversity},
we present how the diversity of given names in each dataset has
evolved over the 20th century and discuss major changes in
diversity. We then summarize this work in Sec.~\ref{sec:summary}.

\section{Datasets}
\label{sec:data}

This work analyzes $10$ family books in Korea, and some of us have
already used them in previous works~\cite{master,kiet,sanghoon}.
From these datasets, we extract the daughters and women married into the
families, and obtain their names and years of birth.
Although the family books cover several centuries,
we obtain a significant number of female names only for the 20th century.
Even in the early 20th century, it was not uncommon for a girl to have no
particular given name.
To have more female names in our dataset, we additionally include
a list of female students enrolled in a university in Korea from 1926 to 1985.
This comprises about 6\% of the number of individuals in our Korean female
dataset.
On the other hand, we exclude male names in the family books from our analysis,
because they are affected too much by a cultural constraint:
Most of Korean male names consist of two syllables and one of them is often
shared by all the male cousins. For example, one of the authors of the present
paper has a name consisting of `Beom' and `Jun',
and his two other brothers Han Jun
and Seong Jun share the latter syllable with him. In this sense, we may regard
only `Beom' as his true identifier, whereas `Jun' is an index for the
generation in the Kim family. Although the total number of male names in the
ten family books is not small, we still find it doubtful that the
statistics is enough to neutralize such a distortion.
In contrast, the brides and the students are sampled from the whole population
and it is hard to imagine any preference for their names. We thus believe that
the set of female names in our data can be regarded as an unbiased sample
of the whole female population in the past.

For comparison, we will also use a dataset of Quebec in
Canada~\cite{Quebecdata1,*Quebecdata2}, in which
the most popular 275 female and 200 male names have been recorded annually
with their frequencies.
In addition, we use another dataset of the United States (US), which
includes all the names that are given to more than five newborn babies every
year~\cite{USdata}.
Due to the one-hundred year time span of the Korean data,
we consider the same period from year 1900 to 2000 for all the others as well.
The number of persons in each dataset is listed as follows:
$342,370$ females in Korea, $1,203,575$ females and $1,205,453$ males in Quebec,
and $163,523,372$ females and $166,237,403$ males in the US, respectively.

\section{Results}
\label{sec:results}

\subsection{Rank-size distribution}
\label{subsec:ranksize}

\begin{figure}
\begin{center}
\includegraphics[width=0.49\textwidth]{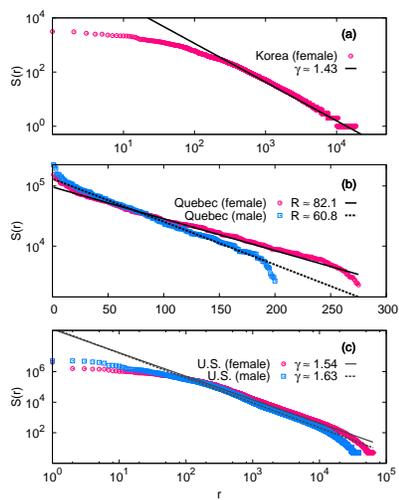}
\end{center}
\caption{Rank-size distribution of given names,
where the vertical axis means the total number of individuals given each
specific name during the 20th century and the horizontal axis means
its corresponding rank.
(a) For Korean females, the rank-size distribution has a fat tail
described as $S(r) \sim r^{-\gamma}$ with $\gamma \approx 1.43$.
(b) The Quebec data exhibit exponential decay such as $S(r) \sim e^{-r/R}$
with a characteristic scale $R \approx 80 (60)$ for females (males).
(c) Similarly to the Korean case, the US data also show broad distribution
with $\gamma \approx 1.54 (1.63)$ for females (males).
}
\label{fig:ranksize}
\end{figure}

We first study rank-size distribution of given names for each dataset.
In most countries except a few,
the rank-size distribution for \textit{family names} is
described as a power law~\cite{master,rossi2013}.
The broadness indicates that it is usually enough
to use family names for distinguishing someone from the others.
The situation is completely different for Korean family names, so they have
an expression roughly translated as `going to Seoul to find someone called
Mr. Kim', which basically means `a needle in a haystack'.
The consideration above naturally leads us to the following idea:
In any human society, the combination of family and given names
will have resolution to distinguish one individual from the others.
If family names already have broad rank-size distribution as in
Western countries, given names do not have to provide further
distinguishability.
On the other hand, if family names have narrow rank-size distribution as
found in Korea, given names must be broadly distributed to make every individual
identifiable. In other words, we expect the role of given names to be
complementary to that of family names.

\begin{table}
\caption{Top 10 most popular names from year $1900$ to $2000$ in
Korea, Quebec, and the US, where F (M) denotes females (males).}
\begin{center}
\begin{tabular}{ c||c|c|c|c|c }
 \hline
 Rank & Korea (F) & Quebec (F) & Quebec (M) & US (F) & US (M)\\
 \hline
1 & Jung Suk & Sylvie & Michel & Mary & James \\
2 & Young Suk & Louise & Pierre & Elizabeth & John \\
3 & Young Ja & Nathalie & Daniel & Patricia & Robert \\
4 & Jung He & Julie & Andr\'{e} & Jennifer & Michael \\
5 & Sun Ja & Diane & \'{E}ric & Linda & William \\
6 & Young He & Chantal & Fran\c{c}ois & Barbara & David \\
7 & Mi Suk & Isabelle & Jean & Margaret & Richard \\
8 & Mi Kyung& Johanne & Claude & Susan & Joseph \\
9 & Kyung He & H\'{e}l\`{e}ne & Martin & Dorothy & Charles \\
10 & Jung Ja & Lise & Alain & Sarah & Thomas \\
 \hline
\end{tabular}
\end{center}
\label{tab:toplist}
\end{table}

Let the size $S$ of a name denote the total number of individuals
given the name during the 20th century. We assign a rank $r$ to each given
name after sorting the data in descending order of $S$
(Table~\ref{tab:toplist}). By construction, $S(r)$ is a non-increasing function.
We indeed find that the rank-size distribution of Korean female names can
be fitted to a power-law form as depicted in Fig.~\ref{fig:ranksize}(a).
It is also consistent with our expectation that given names in Quebec exhibit
exponential rank-size distribution, $S(r) \sim e^{-r/R}$.
Note that the characteristic rank scale $R$ is not known {\it a priori}
from the distinguishability argument.
Our data show that $R \approx 80$ and $R \approx 60$ for females and males,
respectively [see Fig.~\ref{fig:ranksize}(b)].
The larger value of $R$ for female names implies that they are more diverse
than male ones, which will be cross-checked by using a diversity measure below,
but the important point is that they have almost the same order of magnitude
$R \sim O(10^2)$. We suggest that the Dunbar number~\cite{dunbar} could be a
crucial factor to explain this scale:
Suppose that a strongly connected social group, in which
everyone can refer to others on the first-name basis, has a typical size
comparable to the Dunbar number $D$. If $D$ was much greater than $R$,
the first-name basis would be exposed to too much ambiguity. The opposite
limit of $R \gg D$ is again implausible, because we would not need so many
names after all.

The US data show an interesting difference from our expectation, in
that given names are diverse as shown in
Fig.~\ref{fig:ranksize}(c). This is not necessary from our viewpoint,
because family names already provide enough distinguishability~\cite{USdata}.
Our guess is that the fat tail originates from multiethnicity:
As an extremely simple example, suppose that we have mixed the Korean and
Quebec data together. In this mixture, we will find broad
distribution of family names due to the Quebec part, and the given names will
also be broadly distributed because of the Korean part.
Additionally, the interesting relation between Zipf's law and Heaps' law in Ref.~\cite{heaps} 
is not clearly observed in our data.

\subsection{Temporal evolution of popularity of names: Rise and fall}
\label{subsec:riseandfall}

In contrast to family names, given names are not necessarily
inherited, but chosen at parents' disposal.
Although the parents have infinitely many possibilities to choose in principle,
it does not mean that they can choose any: First of all, it should be
acceptable in view of the social norm. For example, New Zealand has banned
disturbing given names such as `Lucifer', `Rogue', and `Mafia'.
It should also be familiar to some extent:
In 1996, a local court in Sweden rejected a name spelled as
`Brfxxccxxmnpcccclllmmnprxvclmnckssqlbb11116'. Due to such social pressure,
the actual choice tends to converge to one of existing names.
However, it would be just pointless if everyone converged to the same choice.
When a name becomes so popular to feel boring, it begins to lose
attractiveness, and another name will take it over.
In a sense, naming a child may be compared to picking out clothes, because
we want the name to be different from others', but not really `out there'.
As an outcome of all this interplay,
we expect a pattern of the rise and fall in popularity,
which is measured by relative frequency in the whole population.

\begin{figure}
\begin{center}
\includegraphics[width=0.49\textwidth]{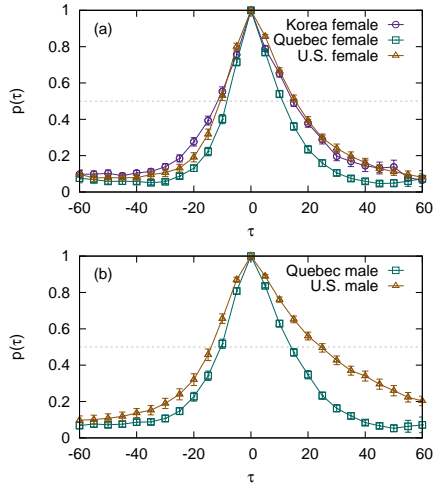}
\end{center}
\caption{Average popularity of top $100$ names in the 20th century.
The time series of each name $i$ is adjusted so as to have
$\tau = 0$ at its maximum and $p_i(\tau=0) = 1$,
after which we take averages over the names (see text for details).
Here, each data point represents popularity for five years.
In most cases,
it roughly takes a decade for the popularity to rise from one half (the
dotted horizontal line) to the maximum. Then, it slowly declines, so
it takes about $15$ years to return back to the half maximum.
Male names in the US have exceptionally slow dynamics,
where the time scales are about $14$ and $25$ years for the rise and fall,
respectively.}
\label{fig:risefall}
\end{figure}

To estimate the popularity of name $i$, we measure its usage fraction $f_i(t)$,
defined as the ratio between the number of newborn babies with $i$
and the total number of newborn babies in year $t$.
We observe that $f_i(t)$ is unimodal for most
names. Only a few names show multiple peaks in $f_i(t)$,
or noisy fluctuations in the time series.
At each $t$, we average $f_i(t)$ over $i$ in the following way:
We first find the peak with height $f_i^{\rm max} \equiv \max_t f_i(t)$,
located at $t_i^{\rm max} \equiv \arg \max_t f_i(t)$.
We then introduce a new variable $\tau \equiv t - t_i^{\rm max}$
so that every name has a peak at $\tau = 0$.
In addition, we define normalized popularity $p_i(\tau) \equiv
f_i(\tau)/f_i^{\rm max}$ so that $p_i(\tau = 0) = 1$ for every $i$.
The range of $\tau$ depends on the value of $t_i^{\rm max}$, hence
is different for each name. Noting that $\tau \in [-100,100]$,
because $t_i^{\rm max}$ is bounded between $1900$ and $2000$,
we restrict the range of $\tau$ to $[-60,60]$
in order to focus on the behavior around $\tau=0$.
Then, we perform `surviving average', which means that we average
$p_i(\tau)$ only for names with nonzero fractions at $\tau$.
The resulting average popularity $p(\tau)$ is plotted
in Fig.~\ref{fig:risefall}.
We notice that the curves in Fig.~\ref{fig:risefall} exhibit
striking similarity: After the initial growth to the half maximum,
it takes about a decade to reach the peak, and about $15$ years
to go back down to one half.
An exception is the US male names, where it takes about twice as long for
popularity to decline, and this seems related to the fact that a boy can
often be named after his father or uncle in the US.
Note that the overall time span corresponds to a couple of generations,
which suggests that popular names in a certain generation would not easily
carry over into the next generation. This observation implies that parents
avoid popular names of their generation when naming their children.
It is also interesting that popularity tends to decline
more slowly compared with the growth, in accordance with Ref.~\cite{fad}.
We quantify the asymmetry between the rise and fall in
Fig.~\ref{fig:risefall}
by measuring the normalized difference between numbers of individuals
before and after $\tau=0$, as listed in Table~\ref{tab:skew}.

\begin{table}
\caption{Asymmetry of the average popularity [$p(\tau)$ in
Fig.~\ref{fig:risefall}] around $\tau = 0$, measured by $[\sum_{\tau > 0}
p(\tau) - \sum_{\tau < 0} p(\tau)]/\sum_{\tau} p(\tau)$, where
F and M denote females and males, respectively.}
\begin{center}
\begin{tabular}{ c | c | c | c | c }
    \hline
    Korea (F) & Quebec (F) & Quebec (M) & US (F) & US (M) \\
    \hline
    $0.10(1)$ & $0.10(1)$ & $0.08(1)$ & $0.16(1)$ & $0.21(1)$ \\
    \hline
\end{tabular}
\end{center}
\label{tab:skew}
\end{table}

\begin{figure*}[ht]
\begin{center}
\center\includegraphics[width=0.8\textwidth]{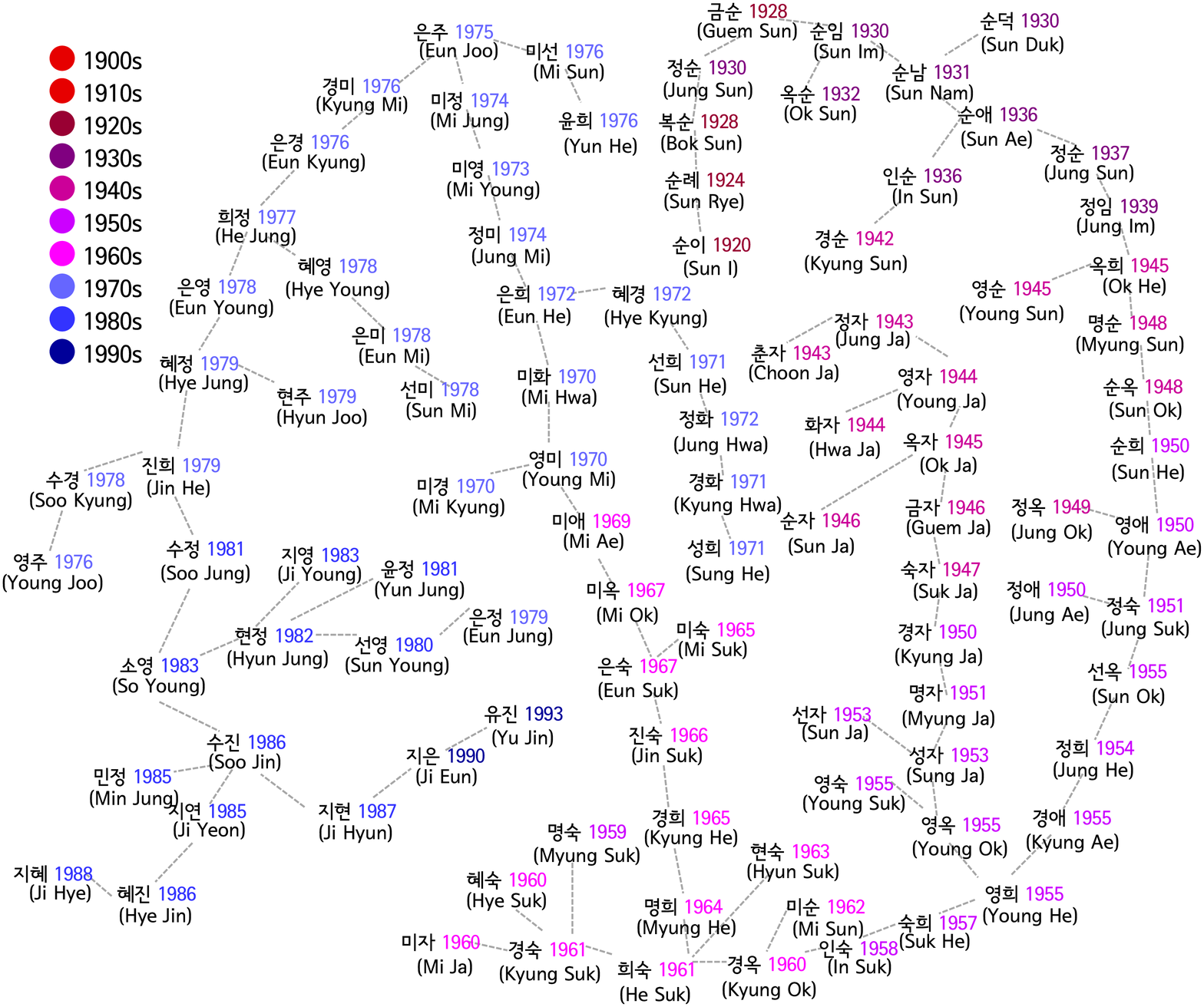}
\end{center}
\caption{MST of Korean female names, where the weight is given by
Eq.~(\ref{eq:dij}). For each name, we show the year when the cumulative
fraction $c_i(t)$ first exceeded $1/2$.}
\label{fig:mstkorea}
\end{figure*}

\begin{figure}[ht]
\begin{center}
\includegraphics[width=0.49\textwidth]{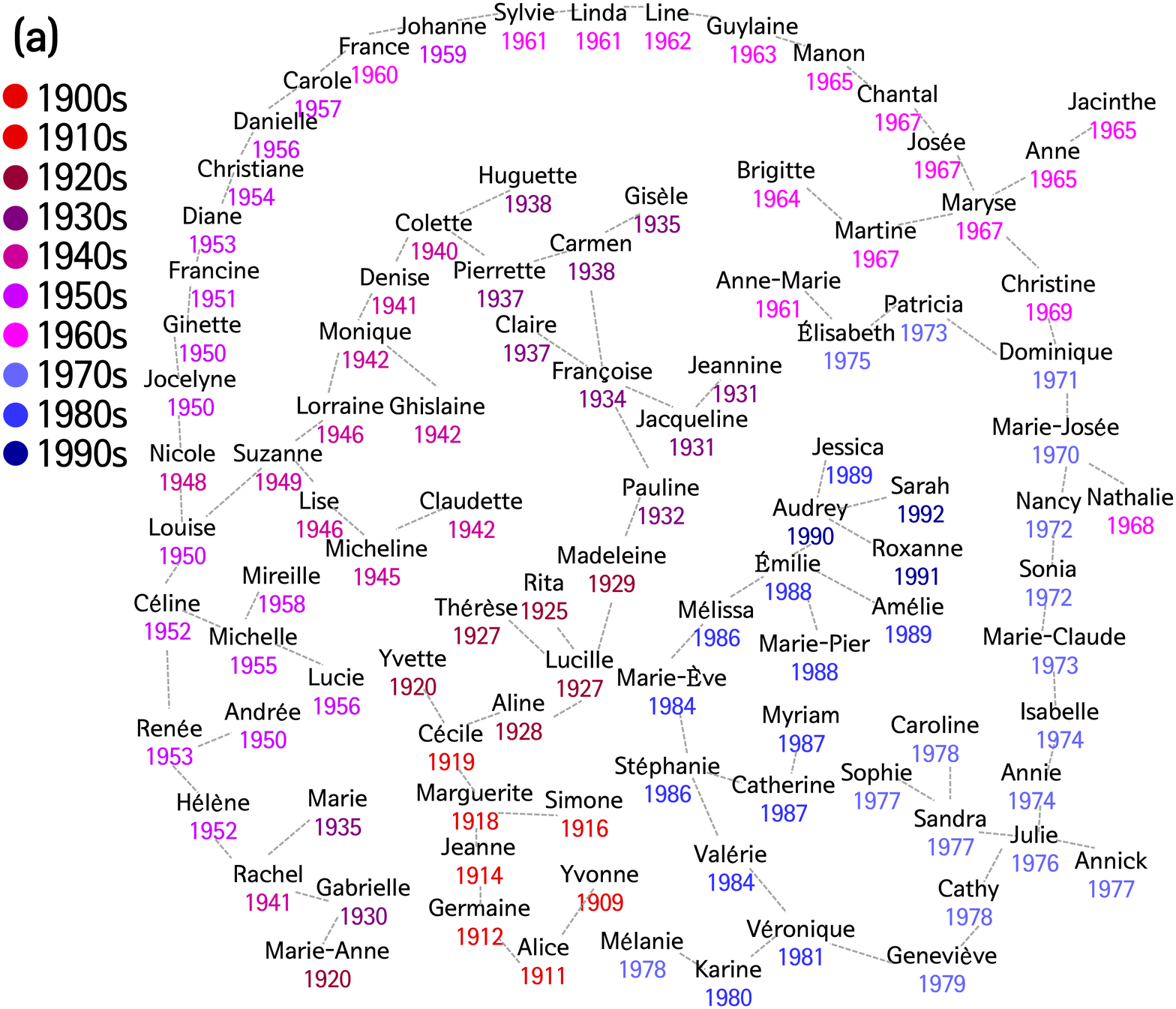}
\includegraphics[width=0.49\textwidth]{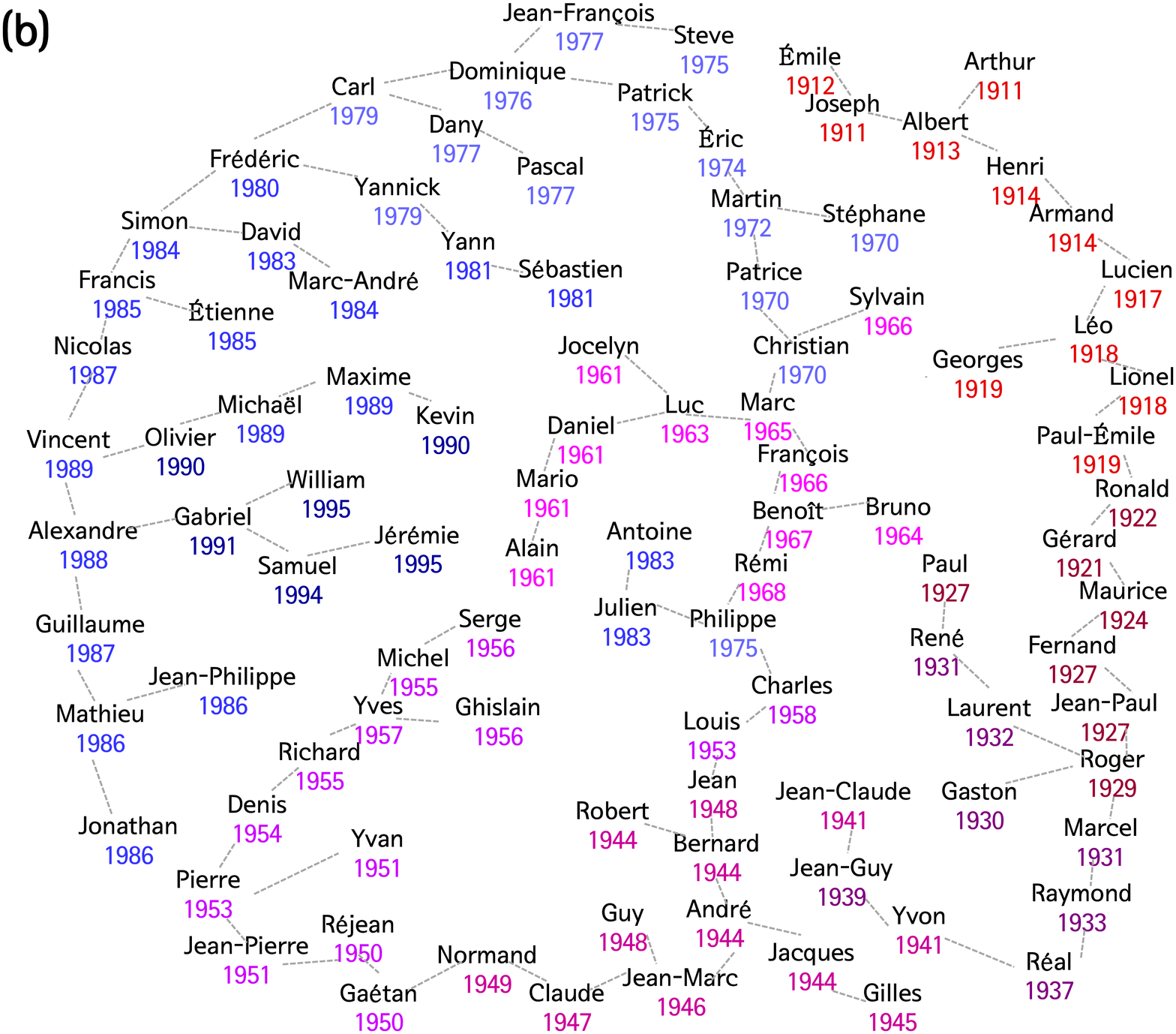}
\includegraphics[width=0.49\textwidth]{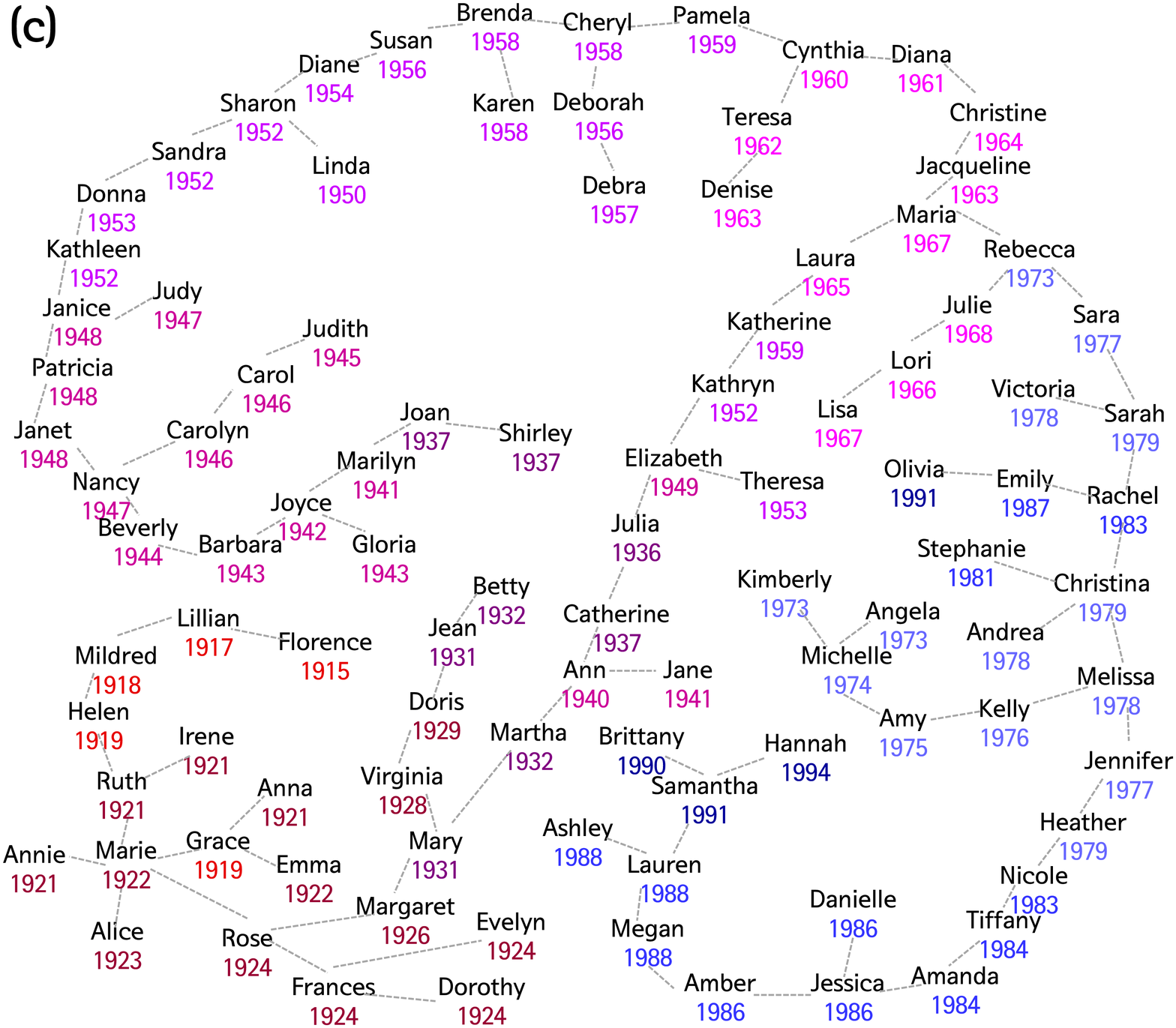}
\includegraphics[width=0.49\textwidth]{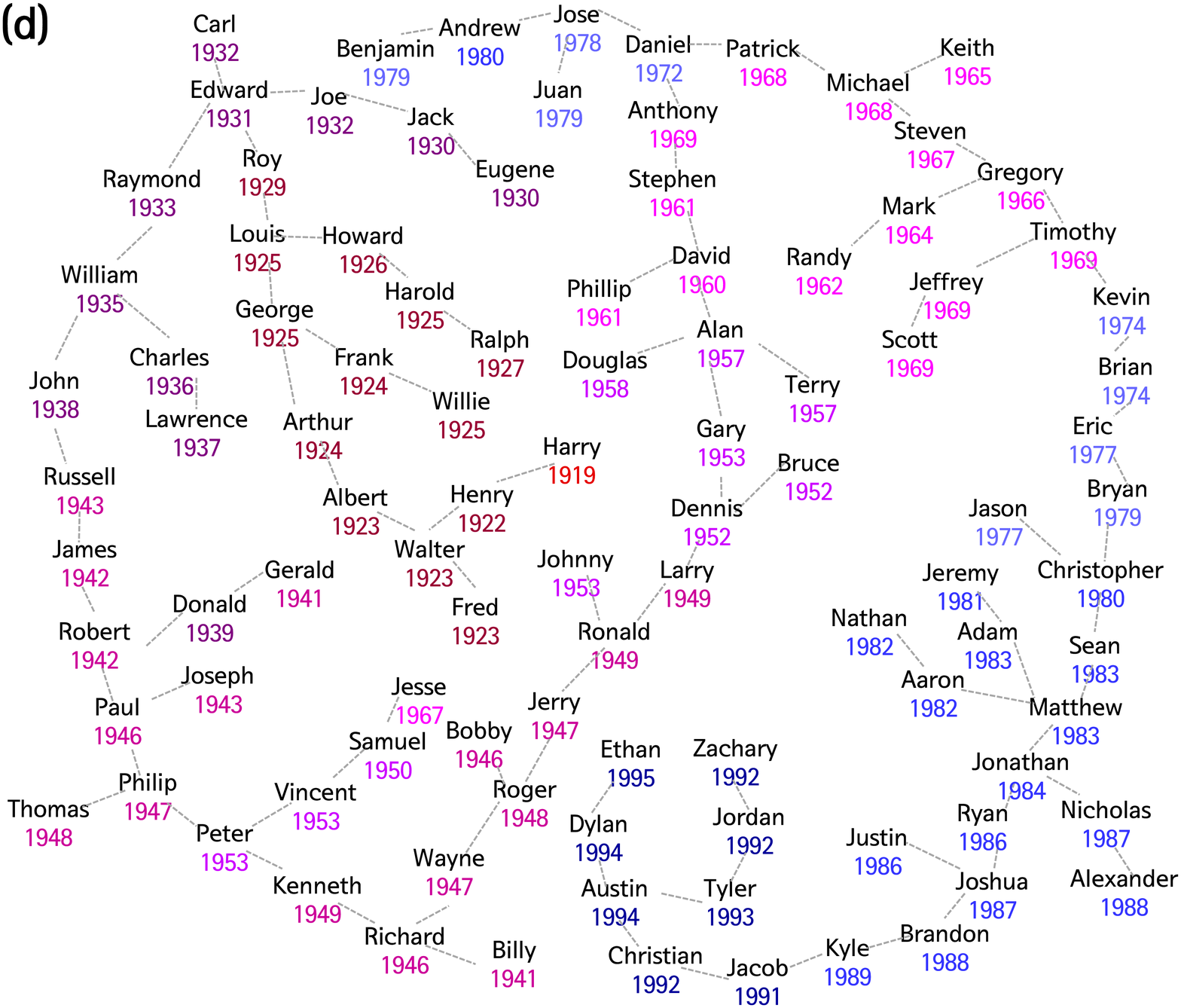}
\end{center}
\caption{MST's for Quebec and the US. The upper panels (a) and (b) show the
cases of female and male names in Quebec, respectively, whereas the bottom
panels are for (c) female and (d) male names in the US.
}
\label{fig:mstothers}
\end{figure}

For better visualization,
we construct a minimum spanning tree (MST) composed of all the hundred names
for each dataset based on the curves $f_i(t)$. To do this, we consider
a cumulative fraction $c_i(t)$ defined as
\begin{equation}\label{eq:cit}
c_i(t)\equiv \frac{\sum_{t'=1900}^{t}f_i(t')}{\sum_{t'=1900}^{2000}f_i(t')}.
\end{equation}
By definition, $c_i(t)$ is a non-decreasing function
which starts from a small value at $t=1900$ and
approaches $c_i(t=2000) = 1$ as $t$ increases.
We then define distance between two names $i$ and $j$ as
\begin{equation}\label{eq:dij}
d_{ij} \equiv \max_t | c_i(t) - c_j(t) |,
\end{equation}
in spirit of the Kolmogorov-Smirnov distance~\cite{ksdistance}.
The idea is that two names, if they are close, will experience similar
time evolution in terms of popularity.
The distance $d_{ij}$ is used as the weight of an edge connecting two
vertices $i$ and $j$. We show the resulting MST's
in Figs.~\ref{fig:mstkorea} and \ref{fig:mstothers}, and
the structure mostly follows the actual chronological order.

\subsection{Diversity}
\label{subsec:diversity}

\begin{figure}[ht]
\begin{center}
\includegraphics[width=0.59\textwidth]{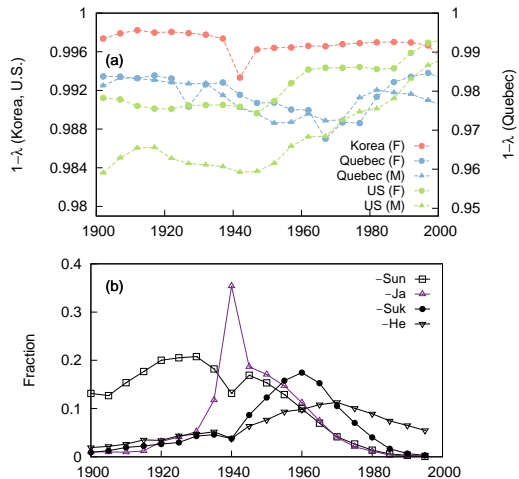}
\end{center}
\caption{(a) Diversity of given names in Korea, Quebec, and the US,
measured by $1-\lambda$, where $\lambda$ is the Simpson index
[Eq.~(\ref{eq:simpson})] and F (M) denotes females (males).
A clear dip in diversity is observed for Korean female names around year 1940,
as a result of the Japanese colonial era (see text for details).
In Quebec and the US, female names tend to be more diverse than male names,
in accordance with the findings in Sec.~\ref{subsec:ranksize}.
(b) Fractions of Korean female names that end with syllables
`Sun', `Ja', `Suk', and `He', respectively.
Names with `Ja' had a sharp peak, recording roughly 40\%, which explains
the dip in diversity in the 1940s.
The errorbars are estimated by reshuffling the data.
}
\label{fig:diversity}
\end{figure}

Our next question is how the diversity of names has changed in time.
In ecology, many diversity indices have been developed to
characterize distribution of species in a biological
domain~\cite{diversity}. Some widely used indices are
the number of species, the Shannon entropy, and the Simpson index.
In this work, we employ the Simpson index $\lambda$, because it is
less sensitive to the total number of species~\cite{diversity}.
Noting that we have names instead of species, we define this quantity as
\begin{equation}\label{eq:simpson}
\lambda(t) \equiv \sum_{i=1}^{N(t)}[f_{i}(t)]^{2},
\end{equation}
where $N(t)$ is the number of names at time $t$, and $f_i(t)$ is as
defined in in Sec.~\ref{subsec:riseandfall}.
This index is closely related to the participation ratio
in the localization problem in quantum mechanics,
where $f_i$ is replaced by probability density $|\psi_i|^2$~\cite{edwards}.
If everyone has a different name,
i.e., $f_i(t) = 1/N$, we have $\lambda = 1/N \ll 1$.
The other extreme is $\lambda = 1$ when everyone has the same name.
To measure diversity, therefore, it is more convenient to look at $1-\lambda$.
Figure~\ref{fig:diversity}(a) shows the time evolution of this diversity
measure for Korean female names.
The biggest change is observed around year 1940, so let us look into this
period in more detail.
Among syllables constituting female names,
we check the most popular four, i.e., `Sun', `Ja', `Suk', and `He',
by collecting names that end with any of these syllables.
We see that names ending with `Ja' had a peak in the early 1940s, decreasing
the diversity, as shown in Fig.~\ref{fig:diversity}(b).
The extensive use of this syllable is traced to a colonial policy:
For the last several years of the Japanese colonial era, between $1940$ and
$1945$, a name-change policy came into effect, forcing
Koreans to change their names to Japanese styles~\cite{soshikamei1,soshikamei2}.
Many Japanese female names ended with a Chinese character meaning a child,
pronounced as `Ko', so the use of this character was the easiest
option for many Korean parents to name their daughters born in the late 1930s
or the early 1940s.
Because the character for meaning a child was pronounced as `Ja' in Korean,
popular Japanese names such as `Yoshiko' and `Junko' became
`Mi Ja' and `Sun Ja', respectively, and these names remained even after
the end of the colonial era.
In Fig.~\ref{fig:diversity}(b), we see that those names with `Ja' once
occupied almost 40\%.
Except for this dip, Korean names have continually exhibited a high degree
of diversity. Differently from our hypothesis,
the diversity does not show appreciable increase in the latter half of
the 20th century, in spite of the change in the status of women during that
period.

The Quebec and US data show very different time series during the same period:
Due to the exponential rank-size distribution, Quebec has low diversity
relative to the other datasets, and male names tend to be less diverse than
female ones. Although the diversity in Quebec had been in gradual decline until
around 1970 and then bounced back, the overall behavior has been quite stable
over the last century in the sense that the diversity in year 2000 is almost
the same as in 1900 with no abrupt changes in between.
In the US, on the other hand, the diversity has been in a long-term uptrend
since the 1950s, which can be explained by the rise in the number of
immigrants~\cite{givenname1, immigrants}.
An interesting point is the plateau in female
names for about two decades from the 1960s. This might be related to the fact
that `Lisa' enjoyed nationwide popularity in the 1960s, as `Jennifer'
did in the 1970s~\cite{USdata}: The former popularity
was boosted by `Lisa Grimaldi' in the soap opera {\it As the World Turns}
launched in 1956,
and the latter is attributed to the great success of {\it Love Story} released
in 1970. Since the 1980s, however, no single given name has swept the entire
US~\cite{USdata}.
This observation suggests that the end of the plateau around 1980 could
signal a transition of such a unipolar state to multipolarity in the cultural
landscape.
This scenario, in turn, provides a way to interpret the plateau in the
Korean case as a marked influence of mass communication, but this
claim calls for more thorough empirical studies.

\section{Summary and conclusion}
\label{sec:summary}
In summary, we have empirically investigated statistics of given names in
Korea, Quebec, and the US.
We have argued that given names play a complementary role to family names
in identifying an individual. In Quebec, for example, it is family names
that work for that purpose, and it is the opposite in Korea. A statistical
consequence of this argument is that if we have limited choices of given
names, family names must diversify, and vice versa. The datasets of Quebec
and Korea have indeed confirmed this prediction.
In the US, both of family and given names are broadly distributed,
and we interpret this observation as a consequence of multiethnicity.
We have also studied how popularity of a name evolves in time, and found
a typical asymmetric pattern of rise and fall with a time scale of
approximately one generation.
As an application of this pattern, we have constructed MST's to visualize
long-term trends of popular given names.
Furthermore, we have suggested the diversity index as a coarse-grained
variable to identify major changes in culture and demography.

Although given-name dynamics is affected by many unpredictable factors,
we conclude that it is also subjected to well-defined constraints,
so that we may expect a striking degree of regularity as long as its
collective patterns are concerned. In a broader context, given-name dynamics
can be understood as a special kind of opinion dynamics in the sense that it
basically
represents opinions of what sounds good and proper as a child's name. Our
finding indicates that an individual's opinion and the surrounding social
pressure may interact in a subtle way: Although they are bound to each other,
it does not mean that one is simply reduced to the other, because it would
mean a loss of individual or social identity. Such a tension yields a
perpetual motion with self-organized patterns in human societies, and the
given-name dynamics gives us fruitful insights into this aspect in a
quantitative manner.

\section*{Acknowledgments}

We are grateful to L. Duchesne for providing us with the data of Quebec and
the US. S.K.B. and B.J.K. were supported by Basic Science Research
Program through the National Research Foundation of Korea
funded by the Ministry of Science, ICT and Future Planning with
grant No. NRF-2014R1A1A1003304 and NRF-2014R1A2A2A01004919, respectively.

\section*{References}

\bibliography{names}

\end{document}